\documentclass[showpacs,aps,prb,twocolumn,epsfig]{revtex4}
\usepackage{amssymb}
\usepackage{graphicx}

\begin{document}

\title{
Combining the advantages of superconducting MgB$_2$ and CaC$_6$ 
in one material: suggestions from first-principles calculations
}
\author{ Amy Y. Liu$^1$}
\author{ I. I. Mazin$^{1,2}$}
\affiliation{$^1$ Department of Physics, Georgetown University, Washington, DC 20057-0995 
\\
$^2$ Center for Computational Materials Science, Naval Research Laboratory,
Washington, DC 20375}

\begin{abstract}
We show that a recently predicted layered phase of lithium monoboride, Li$_2$%
B$_2$, combines the key mechanism for strong electron-phonon coupling in MgB$%
_2$ (i.e., interaction of covalent B $\sigma$ bands with B bond-stretching
modes) with the dominant coupling mechanism in CaC$_6$ (i.e., interaction of
free-electron-like interlayer states with soft intercalant modes). Yet,
surprisingly, the electron-phonon coupling in Li$_{2}$B$_{2}$ is calculated
to be weaker than in either MgB$_2$ or CaC$_6$. We demonstrate that this is
due to the accidental absence of B $\pi$ states at the Fermi level in Li$_{2}
$B$_{2}$. In MgB$_2$, the $\pi$ electrons play an indirect but important
role in strengthening the coupling of $\sigma$ electrons. Doping Li$_{2}$B$%
_{2}$ to restore $\pi $ electrons at the Fermi level is expected to lead to
a new superconductor that could surpass MgB$_{2}$ in $T_c$.
\end{abstract}

\date{\today }
\maketitle

\section{Introduction}

The discovery of superconductivity in MgB$_{2}$ at $T_c =  39$ K led to a new
paradigm for achieving high critical temperatures in conventional
superconductors, namely strong electron-phonon coupling due to metallization
of strong covalent bonds.\cite{AP} In MgB$_2$, this is achieved through the
interaction of electrons forming covalent B-B bonds in the graphene-like
layers with vibrational modes that stretch and compress these 
bonds.\cite{mgb2rev}. Unfortunately, efforts to find related materials 
with even higher 
$T_c$ have not been fruitful. This is basically because the electronic
states that form the covalent bonds in MgB$_2$ are quasi two-dimensional,
which makes it difficult to enhance their density of states and their
coupling to phonons through simple doping.\cite{PickettND} Significant
progress may only be possible through the introduction of different types of
electronic states at the Fermi level.\cite{PickettND}

The recent discovery of superconductivity in CaC$_{6}$ at 
$T_c = 11.5$ K,\cite{cac6} 
the highest critical temperature among graphite intercalation compounds
(GICs), offers an alternative path to increasing $T_c$. Although CaC$_6$ is
structurally similar to MgB$_2$, it appears that the key electrons and
phonons responsible for its superconductivity are distinct from those in MgB$%
_2$.\cite{CM,cac6rev} In particular, free-electron-like states that are
concentrated in the region between graphene layers dominate the
electron-phonon interaction through coupling with soft intercalant modes and
with bond-bending modes of the graphene sheet. Incipient lattice instabilities
associated with the soft intercalant phonon modes, however, may present
obstacles to attaining significantly higher $T_c$ in GICs.\cite{Kim,CM}

It is intriguing to ask whether the MgB$_2$ and CaC$_6$ mechanisms for
strong electron-phonon coupling can be incorporated into a single compound
to create a superconductor with the potential to surpass MgB$_2$. Although
neither MgB$_2$ nor CaC$_6$ can be easily modified so as to incorporate the
key electronic and phonon players from the other material, we show here that
it is indeed possible to combine the MgB$_2$ and CaC$_6$ paradigms in a
different material. Such a material has a good chance of exceeding both MgB$%
_2$ and CaC$_6$ in terms of $T_c$.

The material of interest is a hypothetical layered form of lithium
monoboride, Li$_2$B$_2$, that recent calculations have predicted to be
competitive in energy with known LiB phases.\cite{Curtarolo, Curtarolo2}
The B graphene sheets in Li$_2$B$_2$ are nearly identical to those in MgB$_2$%
. At the Fermi level, there are $\sigma$ bands derived from B $p_{xy}$
orbitals that form covalent bonds within the layers, as in MgB$_2$, and
there are interlayer free-electron-like bands, which we shall call $\zeta$
bands, as in CaC$_6$. By accident, the B $p_z$-derived $\pi$ states that are
present in both MgB$_2$ and CaC$_6$ are missing at the Fermi level in Li$_2$B%
$_2$. Based on the similarity of the in-plane B physics with MgB$_2$,
Kolmogorov and Curtarolo have suggested that this material could be a good
superconductor in the mold of MgB$_2$.\cite{Curtarolo}

In this paper, we analyze the electronic structure of Li$_2$B$_2$ and show
that the electron-phonon interaction has sizable contributions not only from
the $\sigma$ bands that dominate the coupling in MgB$_2$, but also from the
interlayer $\zeta$ bands, which are the key players in CaC$_6$.
Surprisingly, the total electron-phonon coupling in Li$_2$B$_2$ is found to
be weaker than in either of the other two materials. We show that this is
precisely because of the missing $\pi$ electrons at the Fermi level. Though
the $\pi$ bands in MgB$_2$ and CaC$_6$ give only moderate contributions to
the electron-phonon coupling, $\pi$ electrons can indirectly 
strengthen the coupling of $\sigma$ electrons with B bond-stretching 
phonon modes through screening. Doping Li$_{2}$B$_{2}$ to add $\pi $ sheets to
the Fermi surface would enhance the $\sigma$ contribution to the
electron-phonon coupling, as well as add a modest $\pi$ contribution. While
a relatively low $T_c$ is predicted for pristine Li$_2$B$_2$, an
appropriately doped material could have a $T_c$ higher than even MgB$%
_{2}.$

\section{Calculational Methods}

Our analysis is based on density functional calculations. 
The electronic structure was calculated using the WIEN2K general
potential LAPW code,\cite{WIEN} with the GGA functional in the PBE 
form.\cite{PBE} Local orbitals were added to relax the remaining linearization
error and APW orbitals were used to improve the convergence with respect to
the cutoff parameter $RK_{\max }.$ In most calculations $RK_{\max }=7$ was
used, except for the convergence tests with $RK_{\max }=8.$ Up to 918
irreducible \textbf{k}-points were used (32$\times36\times16$). The phonon
spectrum and electron-phonon coupling functions were calculated using the
linear-response method\cite{LR} within the local density approximation.
Norm-conserving pseudopotentials\cite{TM} were used, with a plane-wave
cutoff of 50 Ry. The dynamical matrix was calculated on a grid of $%
6\times6\times2$ phonon wavevectors \textbf{q} in the Brillouin zone and
Fourier interpolated to denser meshes for calculation of the phonon density
of states (DOS). For calculation of the electron-phonon matrix elements, the
electronic states were sampled on grids of $36\times36\times12$ \textbf{k}%
-points in the Brillouin zone.

\section{Results and Discussion}

The structure of the hypothetical layered compound Li$_2$B$_2$, first
proposed in Ref. \onlinecite{Curtarolo} (in which it is called MS2), consists of
honeycomb B sheets, each surrounded above and below by layers of Li on a
triangular lattice. There is a horizontal shift between neighboring Li-B-Li
sandwich units, leading to an ABAB... stacking sequence of the Li-B-Li
trilayer units. While we find that the optimized structural parameters,
particularly the interlayer distances, are sensitive to the details of the
electron-ion potential and to the exchange-correlation functional, the band
structures calculated within the different approximations are very similar.
For the results presented here, we have fixed the structural parameters to
the optimized values reported in Ref. \onlinecite{Curtarolo}.

\begin{figure}[tb]
\includegraphics[width=2.80 in]{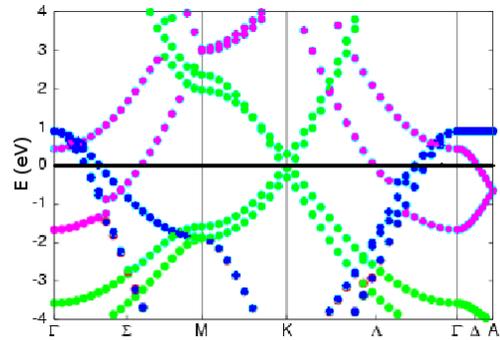} \hspace*{0.03in} %
\caption{(Color online) Li$_2$B$_2$ band structure near the Fermi level
($E_F=0$).
The $\protect\sigma$, $\protect\pi$, and $\protect\zeta$ bands
are plotted with blue, green, and magenta symbols, respectively.}
\label{fig: BS}
\end{figure}

\begin{figure}[tb]
\includegraphics[width=1.4 in]{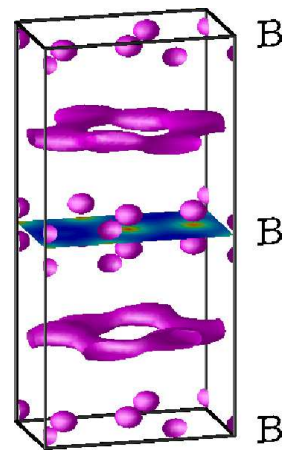}
\caption{(Color online) Isosurface of Li$_2$B$_2$ charge 
density corresponding to  the  $\protect\zeta$ state near the Fermi level, 
half way between the $\Gamma$ and A points.  The top, center, and
bottom plane in the figure contain honeycomb sheets of B. The charge
density in the center B plane is shown. Two Li layers lie between
each pair of B sheets. The $\zeta$ states are  primarily
concentrated in the interlayer region,  but have some B $p_z$ character
as well. 
}
\label{fig: charge}
\end{figure}

The band structure of Li$_{2}$B$_{2}$ near the Fermi level is shown in 
Fig. \ref{fig: BS}. As expected, the $\sigma$ (blue) and $\pi $
(green) states are derived primarily from in-plane B $p_{x,y}$ and
out-of-plane B $p_{z}$ orbitals, respectively. The $\zeta $ bands (magenta)
have substantial Li $sp$ and B $p_{z}$ character. The charge density of the $%
\zeta $ electrons, plotted in Fig.\ref{fig: charge}, is
concentrated in the region between Li-B-Li trilayer units, though
hybridization with B $p_z$ orbitals is evident. This is similar to the
character of $\zeta$ bands in GICs.\cite{Cs,CM,I} The Fermi level in Li$_2$B%
$_2$ lies at the crossing of the $\pi $ and $\pi ^{\ast }$ bands at K,
reminiscent of pure graphite. Note, however, that in graphite, the
coincidence of the Fermi level with the $\pi $-$\pi ^{\ast } $ crossing is
mandated by the absence of other bands near $E_F$, while in Li$_2$B$_2$, it
is accidental. (Expanding $c/a$ by $\sim 10$\%, for example, moves the $\pi$-%
$\pi^*$ crossing away from the Fermi level by more then 0.5 eV.) Compared to
pure graphite, the presence of Li atoms between the B-based graphene sheets
i) makes the $\pi $ bands more 3D and lowers their energy relative to the $%
\sigma $ bands, and ii) lowers the $\zeta $ bands due to the attractive
ionic potential in the interlayer region where the electronic charge in the $%
\zeta $ bands is concentrated.\cite{Cs,cac6rev} With these relative shifts
in the bands, the $\sigma $ bands, which are completely filled in pure
graphite, become partially occupied in Li$_{2}$B$_{2}$, while the $\zeta $
bands, which are unoccupied in graphite, cross the Fermi level in Li$_{2}$B$%
_{2}$. Hence Li$_{2}$B$_{2}$ can be viewed as a self-doped analog to pure
graphite, with holes in the $\sigma $ bands exactly compensated by electrons
in the $\zeta $ band. In contrast, in CaC$_{6}$ the $\pi ^{\ast }$ band is
partially occupied as well.

\begin{figure}[tb]
\includegraphics[width=3.2 in]{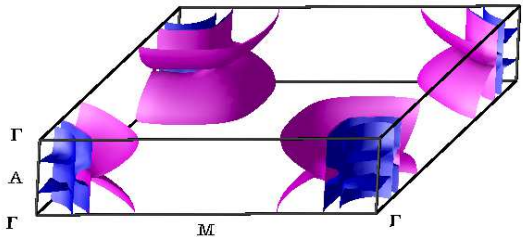}
\caption{(Color online) Fermi surface of Li$_2$B$_2$. The cylindrical sheets 
centered on the $\Gamma$-to-A line arise from the 2D $%
\protect\sigma$ bands, while the ellipsoids are due to free-electron-like 
$\zeta$ bands.}
\label{fig: FS}
\end{figure}

In comparing the band structures of Li$_{2}$B$_{2}$ and MgB$_{2}$, which
both have B honeycomb sheets separated by \textquotedblleft intercalant"
layers that donate one electron per B site, key differences in the relative
energies of the different types of bands can be attributed to the difference
in the separation between B layers. With two Li layers between B sheets in Li%
$_{2}$B$_{2}$, the B-B interlayer distance in Li$_{2}$B$_{2}$ (5.4 \AA ) is
significantly larger than in MgB$_{2}$ (3.5 \AA ). This makes the $p_{xy}$%
-derived $\sigma $ bands more 2D in Li$_{2}$B$_{2}$, but has little effect
on their in-plane dispersion. As in MgB$_{2}$, these bands give rise to
nearly cylindrical hole sheets of the Fermi surface 
surrounding the $\Gamma $ to A line.  As can be seen in the Fermi
surface plotted in Fig. \ref{fig: FS},
there are two nearly degenerate pairs of $\sigma$ cylinders. 
The larger
separation between B planes also causes the B $p_{z}$-derived $\pi $ bands
to be more 2D in Li$_{2}$B$_{2}$ and raises them in energy so that the $\pi
^{\ast }$ band that is partially occupied in MgB$_{2} $ becomes unoccupied
(and just touches $E_F$ at certain points in the Brillouin zone). The Fermi
surface due to the $\pi $ bands thus disappears in Li$_{2}$B$_{2}$. On the
other hand, the free-electron-like $\zeta $ bands, which are unoccupied in
MgB$_{2}$, drop in energy when the interlayer distance is increased since
there is a larger interlayer volume available for the electrons to 
occupy.\cite{cac6rev,note} In Li$_{2}$B$_{2}$, the $\zeta $ bands produce a
slightly squashed ellipsoidal Fermi surface around $\Gamma $, with a smaller
effective mass for motion along the $z$ direction than in the plane. 
These ellipsoids are huge compared to the
small $\zeta$ balls in the CaC$_6$ Fermi surface,\cite{cac6rev} even though 
CaC$_{6}$ has the most occupied $\zeta $ band among all GICs.\cite{Cs}

The topological simplicity of the Fermi surface of MgB$_2$, in which sheets
of the Fermi surface corresponding to bands of different character are
widely separated in reciprocal space, is lost in Li$_2$B$_2$. Not only do
the $\zeta$ ellipsoids cross all four $\sigma$ cylinders, they also cross
each other. At all of these crossings, there is only weak hybridization
between the bands. The multiple crossings between sheets of the Fermi
surface arising from bands of different character, along with the presence
of bands that nearly touch the Fermi level, make Brillouin-zone integrations
sensitive to \textbf{k}-point sampling. This is especially an issue for
calculations of the electron-phonon coupling constants, which involve double
integrals over the Fermi surface. The quantitative electron-phonon coupling
results presented here likely have larger uncertainties than usual but this
does not affect the qualitative picture that emerges.

\begin{figure}[t]
\includegraphics[width=2.5 in ]{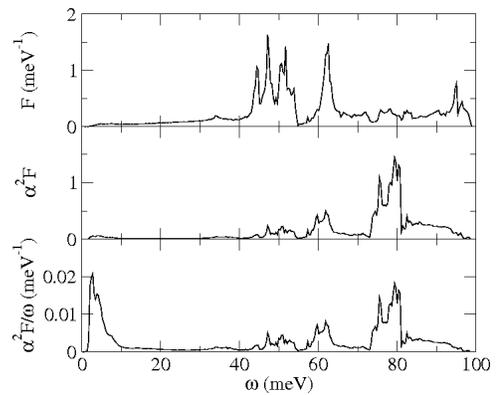}
\caption{Li$_2$B$_2$ phonon density of states $F(\omega)$, 
electron-phonon spectral function $\protect\alpha^2F(\omega)$, 
and ratio of $\protect\alpha^2F$ to frequency $%
\protect\omega$. The electron-phonon coupling constant $\lambda$, which
is proportional to the inverse-frequency-moment of $\alpha^2F$,
has significant contributions from both low-frequency and high-frequency modes.} 
\label{fig: ph}
\end{figure}

The phonon density of states $F(\omega )$ and electron-phonon spectral
function $\alpha ^{2}F(\omega )$ calculated for Li$_{2}$B$_{2}$ are shown in
Fig. \ref{fig: ph}. The $E_{2g}$ in-plane bond-stretching B modes that are
the key players in superconductivity of MgB$_{2}$ are also important in Li$%
_{2}$B$_{2}$, as indicated by the peak in the spectral function near 80 meV.
In both materials, these vibrational modes couple strongly to the B $\sigma $
bands. Given the similarity of the B in-plane physics in the two materials,
one expects similar deformation potentials for the $E_{2g}$ modes.\cite%
{Curtarolo} Indeed, our frozen-phonon calculations using LAPW-GGA produce
exactly the same deformation potential at $\Gamma :$ a displacement $u$
opens a gap $\Delta \varepsilon $ such that $\Delta \varepsilon /u=26$ eV/%
\AA , precisely the same as found for MgB$_{2}$.\cite{AP}
Furthermore, while it is difficult to separate the electronic DOS in Li$_{2}$%
B$_{2}$ into contributions from different bands, as is routinely done in MgB$%
_{2},$ the effective masses derived from the dispersion of the $\sigma $
bands are within 10\% of those in MgB$_{2}$ (the heavy band is 9\% heavier
and the light band 8\% lighter). If the phonon frequencies were the same,
one would then expect the net intraband electron-phonon coupling  constant
for the $\sigma$ bands, $\lambda _{\sigma \sigma }$,
to be practically the same as in MgB$_{2}$. Surprisingly,
the phonon frequencies differ considerably. At the zone center,
the $E_{2g}$ phonon frequency of 81 meV in Li$_{2}$B$_{2}$ is signficantly
higher that the corresponding 67 meV in MgB$_{2}$,\cite{mgb2rev,anh} which
likely reduces $\lambda _{\sigma \sigma }$ from $\approx 1$ to $\approx 0.6
- 0.7.$

The hardening of the in-plane B bond-stretching modes in Li$_{2}$B$_{2}$ is
related to the loss of $\pi $ electrons at the Fermi level.   
The same electron-phonon coupling process  that
gives rise to superconductivity also leads to 
electronic screening of phonons.   For a zone-center phonon 
mode, the softening due to screening by metallic electrons is  
given by
\begin{equation}
\Delta \omega^2 = -4 \omega \langle g^2 \rangle N(0),
\label{eq: softening}
\end{equation}
where $\langle g^2 \rangle $ is a Fermi-surface average of the
square of the electron-phonon matrix element $g$, and $N(0)$ is the 
electronic DOS at the Fermi level.\cite{osv, mgb2rev} 
Because the matrix element $g$ contains a factor of 
$1/\sqrt\omega$, the right-hand side of Eq. \ref{eq: softening} is 
independent of the phonon frequency $\omega$. 
In MgB$_{2}$, both the $\sigma $ and $\pi $ electrons participate in screening 
the bare $E_{2g}$ phonons. 
For the $E_{2g}$ mode in MgB$_2$, it has been estimated that 
$\omega^2$ is reduced by about 1500 meV$^{2}$ due to screening
by $\pi $ electrons, and by about 2000 meV$^{2}$ due to screening by $\sigma 
$ electrons, and that the unscreened frequency is $\omega_0 \approx 90$ 
meV.\cite{mgb2rev} In Li$_{2}$B$_{2}$, the unscreened frequency and the 
softening due to $\sigma $ bands should be roughly the same magnitude as in 
MgB$_2$, which gives a rough estimate for the screened $E_{2g}$ phonon 
frequency of $\omega = \sqrt{\omega_0^{2}-(\Delta\omega^2)_{\sigma} }
\approx 78$ meV. This is consistent with our actual
calculations for Li$_{2}$B$_{2},$ which yield a frequency of 81 meV at 
the zone center. 
Since only phonons with $q_{xy} < 2k_{F}^{\sigma }$, where $k_{F}^{\sigma } $
is the radius of the cylindrical $\sigma $ sheets of the Fermi surface, can
couple to $\sigma$ electrons, softening by $\sigma $ electrons 
is only significant when the in-plane component of the phonon wavevector is 
less than $ 2k_{F}^{\sigma }$.   Indeed, at the zone boundary, 
our calculations yield a frequency of
92 meV for the lowest $E_{2g}$ branch, consistent with
estimates for the unscreened $E_{2g}$ frequency. 
While Li$_{2}$B$_{2}$ also has $\zeta $ electrons, which
are not present in MgB$_{2}$, these are largely localized in the interlayer
region and do not couple strongly with the B bond-stretching modes, so they
are less effective in screening the $E_{2g}$ phonons. 
The harder $E_{2g} $ frequencies in Li$_{2}$B%
$_{2}$ compared to MgB$_{2}$ can thus be attributed to the absence of
screening by $\pi $ electrons.

In addition to the high-frequency $E_{2g}$ peak in the spectral function,
there is a low-frequency peak in $\alpha ^{2}F$ arising from ultrasoft modes
that involve rigid shifts of the B planes in the $z$ direction coupled with
both in-plane and out-of-plane motion of the Li atoms. 
The frequency of these ultrasoft modes
is sensitive to calculational details, so at $\Gamma $, 
for example, the
harmonic frequency of about 8 meV obtained in our LDA linear-response
calculation is somewhat lower than the estimate of 11 meV based on
force-constant methods using PAW GGA calculations.\cite{Curtarolo} In any
case, 
these low-frequency normal modes couple with the $\zeta $ bands, similar to
the situation in CaC$_6$, where Ca sliding modes interact strongly with
interlayer electrons. Although relatively small, the low-frequency peak in $%
\alpha ^{2}F$ gives a similar contribution to the isotropic
electron-phonon coupling constant $\lambda = 2\int d\omega \alpha
^{2}F/\omega $ as 
the high-frequency peak (see Fig. \ref{fig: ph}, bottom
panel). The total $\lambda$ is calculated to be 0.57, with the low-frequency
peak contributing about 0.15, and the $E_{2g}$ peak contributing about 0.2.
The logarithmically averaged phonon frequency of $\omega _{\ln }=\exp [{%
\lambda }^{-1}\int d\omega \ln (\omega )\alpha ^{2}F(\omega )/\omega ]=39$
meV, is well below the average frequency of $\omega _{ave}=59$ meV,
reflecting the importance of the ultrasoft modes.

From the point of view of $T_c$, the prospects for Li$_2$B$_2$ do not look
optimistic compared to MgB$_{2},$ because of both the lack of a $\pi$-band
contribution to $\lambda ,$ and, perhaps more importantly, the hardening of
the key $E_{2g}$ phonons that are no longer screened by $\pi $ electrons.
Neglecting possible multiband effects and using the McMillan
equation\cite{McMillan} 
with a typical Coulomb parameter of $\mu ^{\ast }=0.12, $ we
estimate $T_{c}\approx 7$ K. In comparison, the estimated \textit{isotropic} 
$T_{c}$ in MgB$_{2}$ is 25-30 K. Enhancement of $T_c$ due to interband
anisotropy is expected to be weaker in Li$_{2}$B$_{2}$ since the  
two groups of electrons contribute more equally 
to the coupling with phonons. 
But even if it were of the same order as in MgB$_{2},$ that is, $%
\sim 30\%$ in the effective $\lambda ,$ it would only increase $T_{c}$ to
about 15 K. The bottom line is that Li$_{2}$B$_{2},$ if synthesized, might
be competitive with CaC$_{6}$ in terms of $T_c$, but not with MgB$_{2}.$

However, that is not the end of the story.

While the \textit{undoped} Li$_{2}$B$_{2}$ is isoelectronic with MgB$_{2},$
and doping of MgB$_{2}$ is known to unequivocally depress $T_{c},$ the same
need not be true for Li$_{2}$B$_{2}.$ Electron doping of MgB$_{2}$ 
lowers the $\sigma $ DOS, and hence the electron-phonon coupling,
since the bands are not ideally 2D.  (Though attempts have been
made, hole doping of MgB$_2$ has not yet been achieved.) 
In Li$_{2}$B$_{2},$ 
the $\sigma $ bands are much more 2D, so doping should have
little effect on their DOS and electron-phonon coupling. On the other hand,
electron doping would start filling the $\pi $ bands, increasing
their DOS.  Assuming rigid bands, doping on either B or Li sites
could be used to raise the Fermi level by nearly 1 eV without losing 
the contribution from the $\sigma $ bands.  We estimate that this would 
increase the $\pi$-band DOS to roughly half the value in MgB$_2$.  
Doping on the Li sites (with Al, Be, Mg, etc.) should yield even higher 
$\pi$ DOS, since the extra ionic charge in the Li 
planes would lower the $\pi $ bands relative to the $\sigma $ bands.
Experience shows that it is hard to dope more holes into the $\sigma $ bands
than there already are in pure MgB$_{2}$. Since Li$_{2}$B$_{2}$ starts with 
more holes in the $\sigma $ bands than MgB$_{2}$, it might even be easier to
synthesize electron-doped Li$_{2}$B$_{2}$ than the undoped material.
Assuming that the $\pi $ bands in doped Li$_{2}$B$_{2}$ couple with the $%
E_{2g}$ modes at about the same level as in MgB$_{2},$ this  
additional coupling will soften the $E_{2g}$ phonons and
enhance their coupling with the $\sigma $ bands, without 
strongly affecting the coupling of $\zeta $ bands with soft modes.
This is an extremely favorable combination that is
likely to have a higher $T_{c}$ than MgB$_{2}.$ To a lesser extent the same
holds if the accidental alignment of the Fermi level with the $\pi $-$\pi
^{\ast }$ crossing is modified by uniaxial pressure.

While Li$_{2}$B$_{2}$ and related layered lithium borides proposed 
by Kolmogorov and Curtarolo\cite{Curtarolo} are only 
hypothetical materials as yet, their calculations of the
Li-B phase diagram have suggested a route for synthesizing these materials 
using pressure.\cite{Curtarolo2}  
We have shown that although Li$_{2}$B$_{2}$ seems to
combine the key mechanism for strong electron-phonon coupling
in MgB$_2$ with the dominant coupling mechanism in
CaC$_6$, each of which  has the highest superconducting transition
temperature in its respective class of materials, the transition
temperature of Li$_{2}$B$_{2}$  is expected to be lower 
because of the lack of the seemingly unimportant $\pi$ electrons
at the Fermi level. An  appropriately doped version
of the material, in which $\pi$ electrons are
restored at the Fermi level, however, could result in a promising new 
superconductor that combines the paradigms of MgB$_2$ and CaC$_6$ in
a single material without loss.    Further calculations using 
the virtual crystal approximation  or the supercell approximation
to explore the quantitative effects of doping on the
electron-phonon coupling in this material would be of interest.

\acknowledgments
We thank L. Boeri, G. Bachelet and O.K. Andersen for
useful discussions, and S. Massida and E.K.U. Gross for sharing with us
their work on superconductivity in CaC$_6$ prior to publication. One
of the authors (I.I.M.) thanks Georgetown University for its
hospitality during a sabbatical stay. This work
was supported by NSF Grant No. DMR-0210717.

\end{document}